\documentclass[10pt,twocolumn,english,floatfix,nofootinbib]{revtex4-1}

\usepackage{amsthm}
\usepackage{dsfont}
\usepackage[T1]{fontenc}
\usepackage[latin9]{inputenc}
\usepackage{geometry}
\usepackage{amsmath}
\usepackage{amssymb}
\usepackage{graphicx}
\usepackage{color}
\usepackage{verbatim}

\makeatletter

\usepackage{babel}

\newcommand{\muhat}{\hat{\mu}}

\newcommand{\R}{\mathbb{R}}
\newcommand{\Z}{\mathbb{Z}}

\makeatother

\begin{document}

\title{Gauge-fixing on the Lattice via Orbifolding}

\author{Dhagash Mehta}
\email{dbmehta@ncsu.edu}
\affiliation{Dept. of Mathematics, North Carolina State University, Raleigh, NC 27695, USA,}
\affiliation{Dept. of Chemistry, The University of Cambridge, Lensfield Road, Cambridge CB2 1EW, UK.}
\author{Noah S. Daleo}
\email{nsdaleo@ncsu.edu}
\affiliation{Dept. of Mathematics, North Carolina State University, Raleigh, NC 27695, USA.}
\author{Jonathan D.~Hauenstein}
\email{hauenstein@ncsu.edu}
\affiliation{Dept. of Mathematics, North Carolina State University, Raleigh, NC 27695, USA.}
\author{Christopher Seaton}
\email{seatonc@rhodes.edu}
\affiliation{Dept. of Mathematics and Computer Science, Rhodes College, 2000 N. Parkway Memphis, TN 38112, USA.}

\begin{abstract}
\noindent When fixing a covariant gauge, most popularly the Landau gauge, on the
lattice one encounters the Neuberger 0/0 problem which prevents one
from formulating a Becchi--Rouet--Stora--Tyutin symmetry on the
lattice. Following the interpretation of this problem in terms of Witten-type topological
field theory and using the recently developed Morse theory for orbifolds, we propose a modification of
the lattice Landau gauge via orbifolding of the gauge-fixing group manifold and show that this modification circumvents the orbit-dependence issue and hence can be
a viable candidate for evading the Neuberger problem. Using algebraic geometry,
we also show that though the previously proposed modification of the lattice Landau gauge
via stereographic projection relies on delicate departure from the standard Morse theory due to the non-compactness of the underlying manifold,
the corresponding gauge-fixing partition function turns out to be
orbit independent for all the orbits except in a region of measure zero.
\end{abstract}

\maketitle

\section{Introduction}

\label{sec:intro}
Lattice field theories have proved to be a very successful way of exploring the nonperturbative regime
of quantum field theories. They also provide valuable insight and input to the nonperturbative approaches in the continuum such as
the Dyson-Schwinger equations (DSEs), functional renormalization group studies (FRGs), etc. \cite{Alkofer:2000wg}.
Since each gauge configuration comes with infinitely many equivalent physical copies, the set of which is called
a gauge-orbit, to remove such redundant degrees of freedom from the generating functional, one must fix a gauge
in the continuum approaches.
Hence, to have a direct comparison between the continuum approaches with the corresponding results from the lattice field theories,
one also needs to fix a gauge on the lattice, even though in general gauge-fixing is not required on the lattice due to the manifest gauge invariance
of the lattice field theories. For this reason, gauge-fixed simulations have recently attracted a considerable amount of interest.

In the perturbative limit, the standard approach of fixing a gauge is the Faddeev-Popov (FP) procedure~\cite{Faddeev:1967fc}.
In this procedure, a
gauge-fixing device called the gauge-fixing partition function,
$Z_{GF}$, is formulated out of the gauge-fixing condition. For an ideal gauge-fixing condition, $Z_{GF} =1$.
The unity is then inserted in the measure of the generating
functional, so that the redundant degrees of freedom are removed after
appropriate integration. This procedure was generalized in~\cite{Becchi:1975nq} and is called
Becchi--Rouet--Stora--Tyutin (BRST) formulation. Gribov showed that in non-Abelian gauge theories
a generalized Landau gauge-fixing condition, if treated non-perturbatively,
would have multiple solutions, called Gribov or Gribov--Singer copies~\cite{Gribov:1977wm,Singer:1978dk,Alkofer:2000wg}.
Hence, the effects of Gribov copies should be properly taken into account
within the Faddeev-Popov procedure. In fact, on the lattice, for any Standard Model groups,
the corresponding $Z_GF$ turns out to be zero~\cite{Neuberger:1986vv,Neuberger:1986xz}
due to a perfect cancelation among Gribov copies. Thus, when inserted into the generating functional,
the expectation value of a gauge-fixed observable turns out
to be of the indeterminate form $0/0$, called the Neuberger $0/0$ problem.
The problem yields that a BRST formulation on the
lattice can not be constructed and it is argued this may also hamper comparisons of the results from the lattice
with the continuum approaches~\mbox{\cite{vonSmekal:2007ns,vonSmekal:2008es,vonSmekal:2008ws}}.

In theory, to fix a gauge, one must solve the gauge-fixing condition, a task that could turn out to be extremely difficult
in the nonperturbative regime due to the nonlinearity of the equations. Hence, gauge-fixing is currently formulated as a functional
minimization problem in the lattice field theory simulations because, generally speaking, numerical minimization is a less difficult
task than finding solutions of a system of nonlinear equations.

Let us consider an action that is invariant under the gauge transformation $U_{\bold{j},\mu}\to g_{\bold{j}}^{\dagger}U_{\bold{j},\mu}g_{\bold{j}+\muhat}$,
where $U_{\bold{j},\mu}\in \mbox{SU(}N_C\mbox{)}$ are the gauge-fields, $g_{\bold{j}}\in \mbox{SU(}N_C\mbox{)}$ are the gauge transformations,
$\bold{j}$ is the lattice-site index, and $\mu$ is the directional index. Then,
the standard
choice (using the Wilson formulation of gauge field theories on the lattice) of the lattice Landau
gauge-fixing functional, which we call the na\"ive
lattice Landau gauge functional, to be minimized with respect to
$g_{\bold{j}}$, is
\begin{equation}
F_{U}(g)=\sum_{\bold{j},\mu}\left(1-\frac{1}{N_{c}}\mbox{Re Tr}g_{\bold{j}}^{\dagger}U_{\bold{j},\mu}g_{\bold{j}+\muhat}\right),\label{eq:general_l_g_functional}
\end{equation}
for SU($N_{c}$) groups.
Points which are roots of the first derivatives $f_{\bold{j}}(g):=\frac{\partial F_{U}(g)}{\partial g_{\bold{j}}}=0$
for each lattice site $\bold{j}$ yield the lattice divergence of the lattice
gauge fields and in the na\"ive continuum limit recovers the Landau
gauge $\partial_{\mu}A_{\mu}=0$. The matrix $M_{FP}$ is the
Hessian matrix of $F_{U}(g)$ with respect to the gauge transformations. $Z_{GF}$ is then the sum of the signs of the determinants of $M_{FP}$
computed at the Gribov copies.

The minima of $F_{U}(g)$ are by definition solutions of the gauge-fixing
conditions, but the minima only form a subset of the set of all Gribov copies, since the latter includes
saddles and maxima in addition to the minima.
The set of minima of $F_{U}(g)$ is called the first Gribov region.
There is no cancelation among these Gribov copies, so the Neuberger
$0/0$ problem does not appear if one restricts the gauge-fixing to the space of minima instead of all solutions of the gauge-fixing condition.
This restricted gauge-fixing is called the minimal Landau gauge \cite{Zwanziger:1989mf} and
can be written in terms of a renormalizable action with auxiliary
fields (see, e.g., \cite{Vandersickel:2012tz} for a review).
However, the number of minima may turn out to be different
for different gauge-orbits and increases exponentially with increasing lattice size,
as was shown for the compact U($1$) case in Refs. \cite{Mehta:2009,Mehta:2010pe,Hughes:2012hg,Mehta:2013iea,mehta2014potential,Mehta:2014jla}.
Thus, the corresponding $Z_{GF}$, which counts the number of minima for each gauge-orbit
in the minimal Landau gauge, is orbit-dependent, and inserting $Z_{GF}$ in the generating functional becomes
a difficult task.

To resolve the gauge-dependence issue, one may further
restrict the gauge-fixing to the space of global minima, called the
fundamental modular region (FMR). In this gauge, known as the absolute Landau gauge,
again the Neuberger $0/0$ problem is avoided as in the minimal Landau gauge case.  However,
the corresponding $Z_{GF}$ may be orbit-independent since the number of global minima is
thought to be constant for any gauge-orbit (it is also anticipated that the
set of configurations with degenerate global minima is a set of measure
zero which forms the boundary of the FMR). Thus, the FMR is expected to not have any Gribov copies within it \cite{Zwanziger:1993dh,vanBaal:1997gu}.
This claim was verified to be true for the compact $\mbox{U}(1)$ case for the one- and two-dimensional lattice
in Refs.~ \cite{Mehta:2009,Mehta:2010pe}. The problem with the absolute Landau gauge is that one must
find the global minimum of $F_{U}(g)$
for sampled orbits, which corresponds to finding the global minimum of spin glass model Hamiltonians, a task in most cases known to be an NP
hard problem.

In the past few years, a few further suggestions to evade the Neuberger problem and restore BRST formulations on the lattice have been
put forward in Refs. \cite{Testa:1998az,Kalloniatis:2005if,Ghiotti:2006pm,vonSmekal:2008en,Maas:2009se},
which are reviewed in Ref.~\cite{Maas:2011se}. In the current paper, we concentrate on the stereographic lattice Landau gauge which was proposed in
Refs.~\cite{vonSmekal:2007ns,vonSmekal:2008es,Mehta:2009}. In Section \ref{sec:stereographic_LLG}, we first review this proposed modification
of lattice Landau gauge-fixing via stereographic projection
of the gauge-fixing manifold.  We also give a plausible topological argument on why the proposal might fail. In particular,
the orbit independence of the corresponding $Z_{GF}$ is crucial to ensure that the stereographic lattice Landau gauge is a viable candidate
to evade the Neuberger $0/0$ problem. We also show why topologically the stereographic projection might turn
out to be orbit dependent. In Refs. \cite{Mehta:2009,Mehta:2009zv,Hughes:2012hg}, the problem of finding all Gribov copies on the lattice was transformed
into a problem in algebraic geometry. However, for the stereographic lattice Landau gauge, it was not possible to solve the corresponding equations
using the then available algebraic geometry methods. In Appendix~\ref{app:stereographic_LLG_alg_geom}, with the improved algorithms,
we give explicit calculations
of the number of Gribov copies using an algebraic geometry based method which guarantees to find \textit{all}
isolated solutions for the simplest non-trivial case of the stereographic lattice Landau gauge, i.e., $3\times 3$ lattice with periodic
boundary conditions. With these stronger results, we show that $Z_{GF}$ for the stereographic lattice Landau gauge is orbit independent over
the orbit space except for a region of zero measure.

In Section \ref{sec:orbifold}, we propose a novel modification via orbifolding of the gauge-fixing manifold that is topologically valid unlike the
stereographic case, and show that $Z_{GF}$
is orbit-independent for this gauge-fixing. Though the idea of an orbifold lattice Landau gauge was conceived in 2009 in Ref.~\cite{Mehta:2009},
the necessary mathematical framework, namely, Morse theory for orbifolds, was published later in that year~\cite{Hepworth:2009}. We briefly review
the definition of an orbifold and Morse theory for orbifolds. Then, we apply these concepts to propose a modified lattice Landau gauge based on
orbifolding
of the gauge-fixing group manifold. We show how the modification evades the Neuberger $0/0$ problem for compact U($1$) while maintaining orbit-independence.
We then conclude the paper in Section \ref{sec:conclusions}.

\section{Stereographic Lattice Landau Gauge}
\label{sec:stereographic_LLG}

The following is a review of the stereographic lattice Landau gauge. We start by noting that a
major breakthrough to resolve the Neuberger $0/0$ problem came from Schaden, who in Ref.~\cite{Schaden:1998hz}
interpreted the Neuberger $0/0$ problem in terms of Morse theory.  It can be shown that
the corresponding $Z_{GF}$ for Landau gauge on the lattice
calculates the Euler
characteristic $\chi$ of the group manifold $G$
at each site of the lattice, i.e., for a lattice with $N$ lattice-sites,
\begin{equation}
Z_{GF}=\sum_{\bold{j}}\mbox{sign}(\det\, M_{FP}(g))=(\chi(G))^{N},\label{eq:Z_GF_is_Euler_char}
\end{equation}
where the sum runs over all the Gribov copies. This result is based on
the Poinca\'re--Hopf theorem, which states that the Euler characteristic
$\chi(\mathbb{M})$ of a compact, orientable, smooth manifold $\mathbb{M}$
is equal to the sum of indices of the zeros of a smooth vector field on
$\mathbb{M}$.  In the case that the vector field is the gradient
of a non-degenerate height function, a differentiable function
from the manifold $\mathbb{M}$ to $\mathbb{R}$ with isolated critical points,
the index at a critical point is $\pm 1$ depending on the sign of the Hessian
determinant at the critical point \footnote{It should be emphasised that in Refs.~\cite{Mehta:2009,Mehta:2010pe,Nerattini:2012pi}, it was shown
that the na\"ive lattice Landau gauge is not a Morse function at a few special orbits, such as the trivial orbit, due to the existence of isolated and
continuous singular critical points. However, for a generic random orbit, it is indeed a Morse function and it is this property that saves the
topological interpretation of the gauge-fixing procedure \cite{Schaden:1998hz}.}
From Eq.~(\ref{eq:Z_GF_is_Euler_char}), we identify
$F_{U}(g)$ as
a height function of the gauge-fixing manifold, Gribov copies as the critical points, and $M_{FP}$
as the corresponding Hessian matrix. This interpretation establishes the fact that the gauge-fixing on the lattice
can be viewed as a Witten-type topological field theory \cite{Birmingham:1991ty}.

For compact U($1$), for which the group manifold is $S^{1}$, the link variables and gauge transformations
in terms of angles $\phi_{\bold{j},\mu},\theta_{\bold{j}}\in(-\pi,\pi]\mod2\pi$ are
$U_{\bold{j},\mu}=e^{i\phi_{\bold{j},\mu}}$ and $g_{\bold{j}}=e^{i\theta_{\bold{j}}}$, respectively.
Thus, the na\"ive gauge fixing functional in Eq.~(\ref{eq:general_l_g_functional})
is reduced to
\begin{align}
F_{\phi}(\theta) & =\sum_{\bold{j},\mu}\big(1-\cos(\phi_{\bold{j},\mu}+\theta_{\bold{j}+\muhat}-\theta_{\bold{j}})\big)\nonumber \\
& \equiv\sum_{\bold{j},\mu}(1-\cos\phi_{\bold{j},\mu}^{\theta}),
\label{eq:sllg-functional}
\end{align}
and the corresponding gauge-fixing conditions are:
\begin{equation}
 f_{\bold{j}}(\theta) =-\sum_{\mu=1}^{d}\Big(\sin\phi_{\bold{j},\mu}^{\theta}-\sin\phi_{\bold{j}-\muhat,\mu}^{\theta}\Big)=0,\label{eq:any_dim_sllg_eq}
\end{equation}
where $\phi_{\bold{j},\mu}^{\theta}:=\phi_{\bold{j},\mu}+\theta_{\bold{j}+\muhat}-\theta_{\bold{j}}$.
A given random set of $\phi_{\bold{j},\mu}$ is called a \textit{random
orbit}. Moreover, when all
$\phi_{\bold{j},\mu}$ are zero, it is called the \textit{trivial orbit}.
We choose periodic boundary conditions (PBC) which are given by
$\theta_{\bold{j}+N\muhat}=\theta_{\bold{j}}$ and $\phi_{\bold{j}+N\muhat,\mu}=\phi_{\bold{j},\mu}$,
where $N$ is the total number of lattice sites in the $\mu$-direction.
With PBC, there is a global
degree of freedom leading to a one-parameter family of solutions with $\theta_{\bold{j}}\to\theta_{\bold{j}}+\vartheta,\forall \bold{j}$
where $\vartheta$ is an arbitrary constant angle. We remove this
degree of freedom by fixing one of the variables to be zero, i.e., $\theta_{(N,...,N)}=0$.
Then, $\{\phi_{\bold{j},\mu}\}$ take random values independent of
the action, i.e., the strong coupling limit $\beta=0$, which is sufficient to answer the questions we are interested in this paper.

We can view Eq.~(\ref{eq:sllg-functional})
as a height function from $S^{1}\times\dots\times S^{1}$
to $\mathbb{R}$. Since $\chi(S^{1})=0$, $Z_{GF}=0$.
In fact, for any
compact, connected Lie group $G$ that is not $0$-dimensional,
it is well known that $\chi(G)$ is zero\footnote{To see this,
note that if $t\mapsto g(t)$ is a one-parameter group in $G$ and
$L_{g(t)}$ denotes left-multiplication by $g(t)$, then
the derivative of $L_{g(t)}$ at $t=0$ produces a vector
field on $G$ which never vanishes.  Then $\chi(G)=0$
follows from the Poincar\'{e}--Hopf theorem.}.

To evade the Neuberger $0/0$ problem, Schaden proposed to construct a BRST formulation for the coset space SU($2$)$/$U($1$)
of a SU($2$) theory.  For  this coset space, $\chi$ is non-zero.  The proposal was generalized to fix gauge of
an SU($N_{c}$) gauge theory to the maximal Abelian subgroup
$($U(1)$)^{N_{c}-1}$ in Refs.~\cite{Golterman:2004qv,Golterman:2012ig}. In short, the Neuberger $0/0$ problem
for an SU($N_{c}$)
lattice gauge theory lies in $($U(1)$)^{N_{c}-1}$, and can be avoided
if the problem for compact U($1$) is avoided.

Following this interpretation, a promising proposal to evade the Neuberger $0/0$ problem via
a modification of the gauge-fixing group manifold (i.e., the manifold of the combination
$g_{\bold{j}}^{\dagger}U_{\bold{j},\mu}g_{\bold{j}+\muhat}$)
of compact U($1$) developed using stereographic
projection at each lattice site was presented in Refs.~\cite{vonSmekal:2007ns,vonSmekal:2008es,Mehta:2009}.
The stereographic gauge fixing functional was proposed as:
\begin{align}
F_{\phi}^{s}(\theta) & =-2\sum_{\bold{j},\mu}\ln(\cos(\phi_{\bold{j},\mu}^{\theta}/2)),\label{eq:mllg-functional}
\end{align}
and the corresponding gauge-fixing conditions are:
\begin{equation}
f_{\bold{j}}^{s}(\theta) =-\sum_{\mu=1}^{d}\Big(\tan(\phi_{\bold{j},\mu}^{\theta}/2)-\tan(\phi_{\bold{j}-\muhat,\mu}^{\theta}/2)\Big)=0\label{eq:any_dim_mllg_eq}
\end{equation}
for all lattice sites $\bold{j}$.

Here, the Euler characteristic of the modified manifold is non-zero, so the Neuberger $0/0$ problem is
avoided. Applying the same approach to the maximal Abelian subgroup
(U(1)$)^{N_{c}-1}$, as mentioned above, the generalization as stereographic projection
for SU($N_{c}$) lattice gauge
theories is also possible when the odd-dimensional spheres $S^{2k+1}$,
$k=1,\dots,N_{c}-1$, are stereographically
projected to the real projective space $\mathbb{R}P(2k)$. In those references, using topological arguments
the number of Gribov copies was shown to be exponentially suppressed for the stereographic lattice Landau gauge
compared to the
na\"ive gauge and the corresponding $Z_{GF}$ for the stereographic lattice Landau gauge
was shown to be orbit-independent for compact U($1$) in one dimension.
Since
it can be shown that the FP operator for the stereographic lattice Landau gauge
is generically positive (semi-)definite, $Z_{GF}$ counts the total number of local
and global minima. The stereographic lattice Landau gauge is thought to be
a promising alternative
to the na\"ive lattice Landau gauge, except that the orbit-independence of $Z_{GF}$
was yet to be confirmed for lattices in more than one dimension.

It is interesting to point out that in supersymmetric Yang--Mills
theories on the lattice, non-compact parameterizations of the gauge fields similar to the stereographic
projection have been used \cite{Catterall:2009it}, independently of the development of the
stereographic lattice Landau gauge (see, e.g., \cite{palumbo1990gauge,becchi1992noncompact,becchi1992noncompact}
for earlier accounts on non-compact gauge-fields on the lattice).
The non-compact parameterization in the supersymmetric lattice field theories, unlike the compact (group based)
parameterization, surprisingly avoids the well-known sign problem in these lattice theories \cite{Catterall:2011aa,Galvez:2012sv}.
Recently, a more direct connection between the sign problem
in lattice supersymmetric theories and the Neuberger $0/0$ problem has
been established \cite{Mehta:2011ud} by noticing that the complete action
of, for example, the $\mathcal{N}=2$ supersymmetric Yang-Mills theories in two
dimensions can be shown to be
a gauge-fixing action via Faddeev-Popov
procedure to fix a topological gauge symmetry in this case.

\subsection{Orbit-dependence of the Stereographic Lattice Landau Gauge}

The following provides an explanation of toopologically subtleties of the stereographic gauge (see \cite{Milnor:63,Guillemin:74} for further background).  Let $\mathbb{M}$ be a closed manifold
(i.e., compact and without boundary).
A smooth function
$f:\mathbb{M} \to \R$ has a \emph{critical point} at $x$ if $df_x$ is nonsingular;
a critical point $x$ is \emph{degenerate} if the Hessian $Hf(x)$ of $f$ at $x$ is singular and
\emph{non-degenerate} otherwise.  A \emph{Morse function} is a smooth function whose
critical points are isolated and non-degenerate.  Given such a Morse function of $f$,
the gradient $\nabla f$ is a tangent vector field to $\mathbb{M}$ that vanishes at exactly the critical
points $x \in \mathbb{M}$ for $f$.  As $f$ is Morse, it has isolated critical points, which must
then be finite as $\mathbb{M}$ is closed.  The requirement that a critical point $x$ of $f$ be nondegenerate
implies that the index $\mbox{ind}_x(\nabla f)$ of the vector field $\nabla f$ at
$x$ is $\pm 1$, depending only on the sign of the determinant of the Hessian $Hf(x)$ of $f$ at $x$.
Therefore, letting $C$ denote the set of critical points in $\mathbb{M}$, we have
\begin{align}
\label{eq:SumIndices}
    \sum\limits_{x\in C}\mbox{sign}(\mbox{det}Hf(x))
        &=  \sum\limits_{x\in C} \mbox{ind}_x(\nabla f)     \\
        \nonumber
        &=  \chi(\mathbb{M}),
\end{align}
where the last equality follows from the Poincar\'{e}--Hopf theorem.  Hence,
in the case where $\mathbb{M} = \prod_\bold{j} S^1$ is the product of circles parameterized by the
$\{ \phi_{\bold{j},\mu}^\theta \}$ at each lattice
site, the partition function $Z_{GF}$ in fact depends only on $\mathbb{M}$, and computes
$\chi(\mathbb{M})$ for any collection of $\{\phi_{\bold{j},\mu}\}$ or any choice of Morse function $F_\phi$.

In the case that $\mathbb{M}$ is not closed but rather an open manifold without boundary,
the sum in Eq.~(\ref{eq:SumIndices})
depends on $f$, and not simply on $\mathbb{M}$.  This can be seen,
for instance, by choosing a Morse function on the circle $S^1$ with at least two
critical points (whose indices must sum to $0$) and then by defining $\mathbb{M}$ to be an
open subset of $S^1$.  Then, $\mathbb{M}$ can be chosen to be
an interval in $S^1$ which contains a single critical point~$x$, in which case the
sum is $\pm 1$ depending on $\mbox{ind}_f(x)$.  Also, one can choose
$\mathbb{M}$ to be an open interval in $S^1$ containing no critical points, in which case
the sum is $0$.  Note that in each of these cases, the manifold $\mathbb{M}$ is diffeomorphic
to an open interval. In short, when $\mathbb{M}$ is not closed, the sum of
 the indices depends on the height function.

Using the stereographic gauge fixing functional Eq.~(\ref{eq:mllg-functional}), it can be shown that
the Hessian is generically positive \cite{Hughes:2012hg}, so that $Z_{GF}$ is strictly positive and counts the number
of critical points. For a $1$-dimensional lattice, there are only $N$ critical points \cite{Mehta:2009,vonSmekal:pvtcommun}, so
the corresponding $Z_{GF} = N$, which is independent of orbits, and thus $Z_{GF}$ does not depend on the choice of $\{\phi_{\bold{j},\mu}\}$.
In higher dimensions, however, the above phenomenon may occur, and $Z_{GF}$ may vary with the
choice of $\{\phi_{\bold{j},\mu}\}$ since the stereographic gauge is outside the applicability of Morse theory.

Appendix \ref{app:stereographic_LLG_alg_geom} demonstrates that, for the stereographic lattice Landau gauge for a $2$-dimensional lattice,
the number of Gribov copies and hence $Z_{GF}$ indeed are orbit independent quantities except in a region of orbit space with measure zero,
via explicit calculations.
Specifically, we use an algebraic geometry based method which guarantees to find all isolated solutions of a given
nonlinear system of equations with polynomial-like nonlinearity
to show that though the number of Gribov copies for the $3\times 3$ lattice for the compact U($1$)
case is constant, $11664$, for most of the random orbits $\{\phi_{\bold{j},\mu}\}$, there are regions in the orbit space
for which the numbers of Gribov copies
differ from this number.

\section{Orbifolding}
\label{sec:orbifold}

The following uses orbifolding to develop a modification of lattice Landau gauge which is topologically rigorous unlike the steregraphic gauge.
We start by reviewing some of the basic concepts
about orbifolds. We give the definition of a orbifold and then describe Morse theory for orbifolds. We then apply Morse theory for orbifolds to
propose a modified lattice Landau gauge
via orbifolding the gauge-group manifold that evades the Neuberger $0/0$ problem while being orbit-independent.

Let $\mathbb{M}$ be a manifold and $G$ a finite group of diffeomorphisms of $\mathbb{M}$.
Then the quotient $G\backslash \mathbb{M}$ is an example of a \emph{global quotient orbifold}
or simply \emph{orbifold}.  Note that in general, orbifolds are required to be only
locally of the form $G\backslash \mathbb{M}$, but we restrict our attention here to global
quotient orbifolds; e.g., see \cite{AdemLR:07}.  A point in $G\backslash \mathbb{M}$
corresponds to the $G$-orbit $Gx = \{ gx : g \in G\}$ of $x \in \mathbb{M}$.

There are several Euler characteristics for orbifolds, and each can be computed
using a Morse function with modifications to the method of Eq.~(\ref{eq:SumIndices}).
The reader is warned that the term ``orbifold Euler characteristic'' can refer to
different 
Euler characteristics in the literature.  The most primitive Euler-characteristic, in the
sense that other Euler characteristics can be 
defined in terms of it, is
the so-called \emph{Euler--Satake characteristic} $\chi_{ES}(\mathbb{M}, G)$, which is given by
\begin{equation}
\label{eq:ChiESDef}
    \chi_{ES}(\mathbb{M}, G) = \chi(\mathbb{M})/|G|,
\end{equation}
where $|G|$ denotes the order, or number of elements, of~$G$.
It was defined for general orbifolds in \cite{Satake:1957gb}; see also
\cite{Thurston:1997,BoileauMP:03}.  Note that in general, $\chi_{ES}$ is a rational number.
One may also consider the usual Euler characteristic
(of the underlying topological space) $\chi(G\backslash \mathbb{M})$,
which is related to the Euler--Satake characteristic via
\begin{align}
    \nonumber
    \chi(G\backslash \mathbb{M})
        &=  \frac{1}{|G|} \sum\limits_{g \in G} \chi(\mathbb{M}^g)                               \\
        \label{eq:ChiTopFromES}
        &=  \sum\limits_{(g) \in G_\ast} \chi(\mathbb{M}^g)/|Z(g)|                               \\
        \nonumber
        &=  \sum\limits_{(g) \in G_\ast} \chi_{ES}(\mathbb{M}^g, Z(g)),
\end{align}
where $Z(g) = \{ h\in G : gh=hg\}$,
$\mathbb{M}^g = \{ x\in\mathbb{M} : gx=x\}$ is the set of points
in $\mathbb{M}$ fixed by $g$, $(g) = \{ hgh^{-1} : h\in G\}$ is the
conjugacy class of $g$ in $G$, and $G_\ast$ the set of conjugacy
classes in $G$.  Note that $\chi_{ES}(\mathbb{M}^g, Z(g))$ coincides
for each element of a conjugacy class, so that the last two sums are well-defined.
In particular, $\chi(G\backslash \mathbb{M})$ is the sum of the Euler--Satake characteristics
of the orbifolds $Z(g)\backslash \mathbb{M}^g$, which for $g\neq 1$ are called \emph{twisted sectors}.
The \emph{nontwisted sector} corresponding to $g = 1$ coincides with $G\backslash \mathbb{M}$.
The collection $\sqcup_{(g)\in G_\ast} Z(g)\backslash \mathbb{M}^g$ is called the \emph{inertia orbifold},
denoted $\Lambda(G\backslash \mathbb{M})$, (see e.g. \cite{AdemLR:07}) so that succinctly, the usual
Euler characteristic $\chi(G\backslash \mathbb{M})$
is the Euler--Satake characteristic of the inertia orbifold.

The \emph{stringy orbifold Euler characteristic} $\chi_{str}(\mathbb{M},G)$, introduced in
\cite{Dixon:1985jw,Dixon:1986jc} for global quotients and \cite{Roan:96} for general orbifolds,
see also \cite{AdemR:03}, is defined analogously as
\begin{equation}
\label{eq:ChiSTRDef}
    \chi_{str}(\mathbb{M}, G) = \frac{1}{|G|} \sum\limits_{(g,h) \in G_{com}^2} \chi(\mathbb{M}^{\langle g, h\rangle}),
\end{equation}
where $G_{com}^2$ denotes the set of $(g, h) \in G^2 = G\times G$ such that $gh = hg$
and $\mathbb{M}^{\langle g, h\rangle} = \{x\in\mathbb{M} : gx=hx=x\}$ denotes the set
of points fixed by both $g$ and $h$.
This Euler characteristic is related to the others as follows.
%
For a pair of commuting elements $(g,h)\in G_{com}^2$, let
$[g,h] = \{ (kgk^{-1},khk^{-1}) : k\in G\}$ (the orbit of
$(g,h)$ under the action of $G$ on $G_{com}^2$ by simultaneous
conjugation), let $G_{com\ast}^2 = \{ [g,h] : (g,h)\in G_{com}^2\}$
(the set of orbits), and let $Z(g,h) = Z(g) \cap Z(h)$ denote the
subgroup of $G$ consisting of elements that commute with both
$g$ and $h$.  Then computations similar to those in Eq.~\eqref{eq:ChiTopFromES}
demonstrate that
\begin{align}
    \chi_{str}(\mathbb{M}, G)
        \label{eq:ChiStrFromES}
        &=  \sum\limits_{(g) \in G_\ast} \chi(Z(g)\backslash \mathbb{M}^g)                           \\
        \nonumber
        &=  \sum\limits_{[g, h] \in G_{com\ast}^2} \chi_{ES}(\mathbb{M}^{\langle g, h\rangle},Z(g,h)).
\end{align}
In other words, $\chi_{str}(\mathbb{M}, G)$ is the usual Euler characteristic of the inertia orbifold,
and as well coincides with the Euler--Satake characteristic of the orbifold
$\sqcup_{[g, h] \in G_{com\ast}^2} Z(g,h)\backslash \mathbb{M}^{\langle g, h\rangle}$.  Observe that this latter
disjoint union is in fact the inertia orbifold of the inertia orbifold, which we refer to as the
\emph{double-inertia orbifold} $\Lambda_2(G\backslash \mathbb{M})$.  The orbifold corresponding to
$[g,h]=[1,1]$ is the \emph{nontwisted double-sector}, while the other orbifolds are referred to
as \emph{twisted double-sectors}.  The reader is warned that double-sectors do not coincide with
$2$-multi-sectors defined in \cite{AdemLR:07} unless $G$ is abelian\footnote{The reader may have noticed that the three Euler characteristics $\chi_{ES}$, $\chi$, and $\chi_{str}$
form the $0$th, $1$st, and $2$nd elements of a sequence of Euler characteristics for orbifolds,
so that others can be defined.  This was observed in \cite{AtiyahS:89}, and this sequence was defined
and studied for global quotients in \cite{BryanF:1998}.  More generally, an Euler characteristic
corresponding to each finitely generated discrete group (with the above sequence corresponding to the
groups $\Z^m$ for $m = 0, 1, 2, \ldots$) was assigned to a global quotient an orbifold
in \cite{Tamanoi:01,Tamanoi:03}, and these Euler characteristics were defined for arbitrary orbifolds
in \cite{FarsiS:11}.}.

A \emph{Morse function} on a global quotient orbifold $G\backslash \mathbb{M}$ is 
defined to be a Morse function $f:\mathbb{M}\to\R$ that is \mbox{$G$-invariant}, 
i.e. $f(gx) = x$ for each $g \in G$ and $x \in \mathbb{M}$.  
The latter condition implies that $f$ yields a continuous function
$\tilde{f}: G\backslash \mathbb{M}\to\R$ on the 
topological space $G\backslash \mathbb{M}$ given by $\tilde{f}(Gx) = f(x)$.  
Morse theory has recently been developed for orbifolds in the general
context of Deligne-Mumford stacks \cite{Hepworth:2009} which, in particular, demonstrates that orbifolds always admit Morse functions, and
establishes Morse inequalities for an orbifold and the
\mbox{corresponding~inertia~orbifold}.

To compute the Euler characteristic $\chi_{ES}$ using a Morse function\footnote{Satake worked with \emph{V-manifolds}, orbifolds where each group element is assumed to fix a subset
of codimension at least $2$.  However, this result can be extended to general orbifolds by
applying it to the orientable double-cover, which always satisfies this hypothesis, and can
be proved directly for global quotient orbifolds using the Poincar\'{e}--Hopf theorem for
manifolds.}, one can apply the
Poincar\'{e}--Hopf theorem for orbifolds as demonstrated in Ref. \cite{Satake:1957gb}.

For a global quotient orbifold $G\backslash \mathbb{M}$, a point $Gx$ is a
\emph{critical point} of $\tilde{f}$ if $x$ is a critical point of $f$, and $Gx$ is said
to be \emph{degenerate} (respectively \emph{non-degenerate}) if $x$ is
degenerate (respectively non-degenerate) for $f$.  Note that
the requirement that $f$ is $G$-invariant implies that these notions
do not depend on the choice of $x$ from the orbit $Gx$.

Similarly, the gradient $\nabla f$
(depending on a choice of Riemannian metric)
defines a $G$-equivariant vector field on $\mathbb{M}$,
which induces a vector field denoted $\nabla \tilde{f}$ on the orbifold $G\backslash \mathbb{M}$.  If
$Gx$ is a zero of $\nabla \tilde{f}$ (equivalently, a critical point of $\tilde{f}$), then the
index of $\nabla \tilde{f}$ at $Gx$ is defined to be
\[
    \mbox{ind}_{Gx}^{orb}(\nabla \tilde{f}) =   \frac{1}{|G_x|}\mbox{ind}_x(f)
\]
where $G_x=\{ g\in G: gx=x\}$ is the subgroup of $G$ that fixes $x$.
That is, the index of a critical
point on an orbifold is the index of a corresponding critical point on the manifold
divided by 
$|G_x|$.  Again, the (manifold) index
can be computed as the sign of the determinant of the Hessian.

If $C$ denotes the set of critical points of $\tilde{f}$ on $G\backslash \mathbb{M}$, then
Satake's Poincar\'{e}--Hopf theorem for orbifolds implies
\begin{align*}
    \sum\limits_{Gx\in C} \frac{1}{|G_x|} \mbox{sign}(\mbox{det} Hf(x))
        &=  \sum\limits_{Gx\in C} \mbox{ind}_{Gx}^{orb}(\tilde{f})          \\
        &=  \chi_{ES}(\mathbb{M}, G).
\end{align*}
Therefore, the sum of the (orbifold) indices of the critical points computes the Euler--Satake
characteristic.  In the context of global quotients, it is not hard to show that a Morse function
$\tilde{f}$ on $G\backslash \mathbb{M}$ defines a Morse function $\Lambda \tilde{f}$ on the inertia orbifold
$\Lambda(G\backslash \mathbb{M})$ as well as a Morse function $\Lambda_2 \tilde{f}$ on the double-inertia
orbifold $\Lambda_2(G\backslash \mathbb{M})$ by restricting $\tilde{f}$ to the appropriate fixed-point
submanifolds.  By Eq.~(\ref{eq:ChiTopFromES}) and (\ref{eq:ChiStrFromES}), we obtain that
applying the procedure above to $\Lambda \tilde{f}$ or $\Lambda_2 \tilde{f}$ yields
$\chi(G\backslash \mathbb{M})$ and $\chi_{str}(\mathbb{M},G)$, respectively.

\subsection{A simple example}

To illustrate this procedure, consider the orbifold given by $\mathbb{M} = S^1$ and $G = \Z_2$,
where the nontrivial element $a$ of $\Z_2$ acts via $e^{i \theta} \mapsto e^{-i\theta}$.
The resulting orbifold
can be identified with $\{ e^{i\theta} : 0\leq \theta\leq\pi \}$, as each $e^{i\theta}$
with $\pi<\theta<2\pi$ is in the orbit of $e^{i(2\pi-\theta)}$.
It is therefore homeomorphic to a closed interval, where the endpoints
are the images of the two points fixed by $\Z_2$.  Then we have that
$\chi_{ES}(\mathbb{M},G) = 0$, as $\chi(S^1)=0$, and
$\chi(G\backslash \mathbb{M}) = 1$, the Euler characteristic of a closed interval.  To compute
$\chi_{str}(\mathbb{M},G)$, note that all elements of
$G^2=\{(1,1),(1,a),(a,1),(a,a)\}$ are mutually commuting,
and the common fixed-point set of each is two points except for the trivial pair
$(1, 1)$ which fixes all of $S^1$.  Hence, applying Eq.~(\ref{eq:ChiSTRDef}) yields
$\chi_{str}(\mathbb{M},G) = 3$.

To compute these Euler characteristics using a Morse function, we choose
$f(\theta) = \cos(\theta)$.  The corresponding $\tilde{f}$ has critical points
at the orbits of $\theta = 0$ and $\theta = \pi$.  The Hessians of $f$
at these two critical points are $-1$ and $1$, respectively, and the isotropy
groups are both $\Z_2$, so that we compute
\begin{align*}
    \chi_{ES}(\mathbb{M},G)
        &=  \mbox{ind}_{G0}^{orb} (\nabla \tilde{f})
            + \mbox{ind}_{G\pi}^{orb} (\nabla \tilde{f})            \\
        &=  \frac{-1}{|\Z_2|} + \frac{1}{|\Z_2|}
        \quad = \quad - \frac{1}{2} + \frac{1}{2} =  0.
\end{align*}

To compute $\chi$, we note that the inertia orbifold $\Lambda(G\backslash \mathbb{M})$ in this case
has three connected components,  the nontwisted sector as well as two points, each equipped
with trivial $\Z_2$-actions.  The function $f$ restricted to a point trivially
has a non-degenerate critical point of index $1$.
It follows that $\chi(G\backslash \mathbb{M})$ is given by the sum of
$\chi_{ES}(\mathbb{M},G)$, computed above, as well as
one term of the form $1/|\Z_2| = 1/2$ for each twisted sector.
That is,
\[
    \chi(G\backslash \mathbb{M})
    =
    \chi_{ES}(\mathbb{M},G) + \frac{1}{|\Z_2|} + \frac{1}{|\Z_2|} = 1.
\]
Similarly, as $\Lambda_2(G\backslash \mathbb{M})$ consists of $G\backslash \mathbb{M}$ as well as six points,
each with isotropy $\Z_2$, we have
\[
    \chi_{str}(\mathbb{M},G)
    =
    \chi_{ES}(\mathbb{M},G) + 6\left(\frac{1}{|\Z_2|}\right) = 3.
\]

\subsection{Orbifolding the lattice Landau gauge}

To apply the lattice Landau gauge procedure for compact $\mathrm{U}(1)$ to orbifolds,
we define a $\Z_2$-action on the space variables $\{ \phi_{\bold{j},\mu}^\theta \}$
by letting the nontrivial element $a \in \Z_2$ act via
$a: \phi_{\bold{j},\mu}^\theta\mapsto -\phi_{\bold{j},\mu}^\theta$.
The choice of group action is motivated by the fact that the
gauge fixing function Eq.~(\ref{eq:sllg-functional}) is invariant under this action.
However, though it is the case that $\chi_{ES}((S^1)^{N^d-1},\Z_2) = 0$,
neither $\chi(\Z_2\backslash(S^1)^{N^d-1})$ nor $\chi_{str}((S^1)^{N^d-1},\Z_2)$
vanish.  The inertia orbifold $\Lambda(\Z_2\backslash (S^1)^{N^d-1})$
consists of the nontwisted sector as well  as $2^{N^d-1}$ points with
trivial $\Z_2$-action, each given by the orbit of a point
$(\phi_{\bold{j},\mu}^{\theta})$ where each $\phi_{\bold{j},\mu}^{\theta}$ is $0$ or $\pi$,
so that
\[
    \chi\big(\Z_2\backslash (S^1)^{N^d-1}\big)
        =   2^{N^d-2}.
\]
Similarly, as each of the pairs of group elements $(1, a)$, $(a, 1)$, and $(a,a)$
fix again $2^{N^d-1}$ points, the double-inertia $\Lambda_2 (\Z_2\backslash (S^1)^{N^d-1})$
consists of the nontwisted sector and $3\cdot 2^{N^d-1}$ points with trivial $\Z_2$-action,
so that
\[
    \chi_{str}\big((S^1)^{N^d},\Z_2\big)
        =   3\cdot 2^{N^d-2}.
\]

To apply the procedure, then, given a random choice of $\{ \phi_{\bold{j},\mu}\}$,
is to use the Morse function $\tilde{F}$ on $\Z_2\backslash (S^1)^{N^d-1}$
induced by $F$ on $(S^1)^{N^d-1}$ defined in Eq.~(\ref{eq:sllg-functional})
with no changes to the  gauge-fixing and boundary conditions.  Since
$\Lambda_2(\Z_2\backslash (S^1)^{N^d-1})$ consists only of the nontwisted
double-sector and $0$-dimensional twisted double-sectors,
the restriction of $\Lambda_2\tilde{F}$ to each connected component of a twisted double-sectors
trivially has a nondegenerate critical point with positive index.  Hence, if
$C$ denotes the set of critical points on the nontwisted sector, we have
\[
    Z_{GF}  =   \sum\limits_{\Z_2\theta \in C} \frac{1}{|(\Z_2)_\theta|}\mbox{sign}(\mbox{det} M_{FP})
                + 3\cdot 2^{N^d-2} = 3\cdot 2^{N^d-2}.
\]
Note that the sum vanishes because it computes $\chi_{ES}((S^1)^{N^d-1},\Z_2) = 0$.
Hence the critical points in the nontwisted sectors occur in pairs with positive and negative
Hessian determinants.
Furthermore, note that the computation of the sum differs from the manifold case in that
a pair of stationary points $\{\phi^\theta_{\bold{j},\mu}\}$ and $\{-\phi^\theta_{\bold{j},\mu}\}$ of $F$ are the same stationary point
for $\tilde{F}$, and hence the sign of $\mbox{det}M_{FP}$ is counted only once.
This may be accomplished algebraically by choosing a single
$\phi^\theta_{\bold{j},\mu}$ and considering only critical points such that $0\leq\phi^\theta_{\bold{j},\mu}\leq\pi$; for critical points such that
 $\phi^\theta_{\bold{j},\mu} = 0$
or $\pi$, we choose another variable and restrict in the same way.

As an example, let $N=3$ and $d=1$.  We consider the trivial
orbit for simplicity, i.e. each $\phi_i = 0$, and fix $\theta_3 = 0$ to remove
the global degree of freedom arising from the periodic boundary condition $\theta_{N+3} = \theta_i$; see Section \ref{sec:stereographic_LLG}.
Then we have
\begin{align*}
    F_{\phi}(\theta)
        &=  \sum_{i=1}^{N} (1-\cos \phi^{\theta}_{i})
        \\&=    3-\cos (\theta_{2}-\theta_{1})
                -\cos (-\theta_{2})
                -\cos (\theta_{1}).
\end{align*}
Setting $\frac{\partial}{\partial \theta_i}F_{\phi}(\theta)=0$
for $i=1,2$, we find five solutions for $(\theta_1,\theta_2)$:
$(0,0)$, $(0,\pi)$, $(\pi,0)$, $(\pi,\pi)$, and $(2\pi/3, -2\pi/3)$.
Note that we only
consider solutions such that $0 \leq \theta_1 \leq \pi$ as above,
because the solution $(-2\pi/3, 2\pi/3)$
is in the same $\Z_2$-orbit as $(2\pi/3, -2\pi/3)$ and hence represents
the same point on the orbifold.  The Hessian determinants of these
critical points are $+3, -1, -1, -1$, and $3/4$, respectively,
and the first four critical points are fixed by
$\Z_2$ while the last is fixed only by the trivial element.
It follows that the indices are given by $1/2, -1/2, -1/2, -1/2$, and $1$,
respectively, and their sum computes $\chi_{ES}\big((S^1)^2,\Z_2)=0$.
To compute $\chi\big(\Z_2\backslash(S^1)^2)$, we consider
$F_{\phi}(\theta)$ as a function on the larger space
$\Lambda\big(\Z_2\backslash(S^1)^2)$ consisting of
$\Z_2\backslash(S^1)^2$ as well as four isolated points fixed
by $\Z_2$ corresponding to the fixed points $(0,0)$, $(0,\pi)$, $(\pi,0)$,
and $(\pi,\pi)$.  Each point is isolated and hence trivially a critical
point with index $1/|\Z_2|$, so summing these indices along with those
on $\Z_2\backslash(S^1)^2$ described above yields
$\chi\big(\Z_2\backslash(S^1)^2) = 2$.  For
$\Lambda_2\big(\Z_2\backslash(S^1)^2)$, we consider instead three copies of each
isolated fixed point, one for each nontrivial commuting pair $(1,a)$, $(a,1)$,
and $(a,a)$, yielding twelve critical points with index $1/|\Z_2|$ and hence
$\chi_{str}\big((S^1)^2,\Z_2) = 6$.

\subsection{An Integral Formulation of $Z_{GF}$ for Orbifolding}
For the sake of completeness, we also provide an expression of $Z_{GF}$ in the usual
integral formulation a la Faddeev-Popov procedure, which we
plan to further refine to suit the needs of the lattice simulations.
To compute the topological Euler characteristic $\chi(\Z_2\backslash(S^1)^{N^d-1})$,
we have
\begin{equation}
 Z_{GF} =   \int_{\Lambda(\Z_2\backslash(S^1)^{N^d-1})}^{orb}
            \mathcal{D}\theta \mathcal{D}\phi
            \prod_{i=1}^{N^d-1}\delta(f_{i}) \frac{H(F)}{|H(F)|}
\end{equation}
where $\mathcal{D}\theta$ indicates integration over each $\theta$, the $f_{i}$
are the stationary equations, i.e., $f_{i} = \frac{\partial F}{\partial \theta_{i}}$,
and $H(F)$ is the hessian determinant of $F$.  The integral is computed in the orbifold sense,
see \cite[Section 2.1]{AdemLR:07}.  If we let $X$ denote the subset of
$(S^1)^{N^d-1} \times \Z_2$ consisting of pairs $(\theta,g)$ such that $g\theta=\theta$,
then this orbifold integral can be expressed using the usual integral as
\begin{equation}
 Z_{GF} =   \frac{1}{2} \int_{X} \mathcal{D}\theta \mathcal{D}\phi
            \prod_{i=1}^{N^d-1}\delta(f_{i}) \frac{H(F)}{|H(F)|},
\end{equation}
where the prefactor $1/2$ arises from the order of $\Z_2$ and the definition of orbifold
integration.

\subsection{Summary of the Procedure}
To summarize, the procedure for computing the topological and stringy Euler characteristic
from the naive gauge-fixing functional can be divided in three steps.
In the first step:
\begin{enumerate}
 \item Find all the stationary points of $F_{\phi}(\theta)$ as given in Eq.~\ref{eq:sllg-functional}
 by solving $\frac{\partial F}{\partial \theta_i} = 0, i=1,...,N^d$.
\item Call the solution vectors of these equations $\phi^{\theta}$.  Let's say there are $M$ solutions.
\item If for two solutions, say $\phi^{\theta (1)}$ and $\phi^{\theta (2)}$, we have $\phi^{\theta (1)} = - \phi^{\theta (2)}$,
then discard one of them. Hence, $m\leq M$ solutions are left in the end.
\item Compute the hessian determinant at each of the $m$ solutions.
\item For each solution $\phi^{\theta}$, the index is $\pm 1$ if $\phi^{\theta} \neq -\phi^{\theta}$
and $\pm \frac{1}{2}$ if $\phi^{\theta} = -\phi^{\theta}$, where the sign is that of the hessian determinant.
\item Compute the sum of the indices for each solution. This sum will be always zero in our case.
\end{enumerate}

For the second step (the fixed points):
\begin{enumerate}
\item The fixed vectors are simply $\phi^{\theta} = {(0,0,...,0), (0,0,...,0,\pi),...,(\pi, \pi,...,\pi)}$, i.e., all the $2^{N^d-1}$
combinations of $0$ and $\pi$.  These solutions already appeared in the first step, but are now considered
as isolated points (twisted sectors) associated to the nontrivial group element.
\item By convention, the `hessian determinant' for each of these solutions is positive, and each solution is fixed by construction, so the index of each of these points is $+\frac{1}{2}$.
\item The (topological) Euler characteristic $\chi(G\backslash \mathbb{M})$ is given by the sum of all indices found
in the first two steps, $\chi(G\backslash \mathbb{M}) = 0 + (\frac{1}{2})\cdot 2^{N^d-1} = 2^{N^d-2}$.
\end{enumerate}

Finally, the third step (for the fixed points associated to commuting pairs):
\begin{enumerate}
\item The fixed vectors are the same as in the second step, but we now consider three copies of each
        for the three nontrivial commuting pairs of group 
        elements~($(a,1)$,~$(1,a)$,~and~$(a,a)$~where~$a$~is~the~nontrivial~element~of~$\Z_2$).
\item We again have that the index of each such point~is~$+\frac{1}{2}$.
\item The stringy Euler characteristic of the orbifold is then the sum of the indices from first and
third step, i.e.,
$Z_{GF} = \chi_{str}(\mathbb{M}, \mathbb{Z}_2) = 0 + (\frac{1}{2})\cdot 3\cdot 2^{N^d-1} = 3\cdot 2^{N^d-2}$.
\end{enumerate}

\section{Conclusion}
\label{sec:conclusions}

Like many other crucial nonperturbative phenomena, gauge-fixing and the BRST symmetry are yet to be well understood in the nonperturbative
regime of gauge field theories. In this paper, we first reviewed and investigated a recently proposed modified Landau gauge on the lattice, known as stereographic lattice Landau gauge.
We gave plausible arguments to demonstrate why this gauge may not turn out to be a valid topological field theory due to the fact that the procedure
is outside the applicability of Morse theory. In Appendix~\ref{app:stereographic_LLG_alg_geom}, we use algebraic geometry to
show for the simplest non-trivial example of $3\times 3$ lattice with periodic boundary conditions for compact U($1$)
that though the number of Gribov copies for the stereographic lattice Landau gauge remains constant for almost all the random gauge-orbits, there are
certain regions in the gauge-orbit space for which the number of Gribov copies differs from the generic case. Since the corresponding $Z_{GF}$ counts
the number of Gribov copies for the stereographic lattice Landau gauge, our results yields that the $Z_{GF}$ is orbit independent over the orbit space
except for a
region with measure zero.

We then proposed modified lattice Landau gauge via orbifolding of the gauge-fixing manifold which is mathematically more rigorous due to the recently
developed Morse theory for orbifolds. We reviewed the definition and description of Morse theory
for an orbifold. We also discussed three different Euler characteristics of an orbifold. We then demonstrated how Morse theory for orbifolds can be
applied to modify the na\"ive lattice Landau gauge so that the corresponding
$Z_{GF}$ for the orbifold lattice Landau gauge, which computes the stringy (or the usual) Euler characteristic of an orbifold,
is orbit-independent and also evades the Neuberger $0/0$ problem since the Euler characteristic is non-zero.
The orbifolds we considered are always compact since the original manifold is compact.
Thus, our modified lattice Landau gauge is fundamentally different
than the stereographic lattice Landau gauge in that the former retains the compactness of the gauge-fixing manifold, and is close in the spirit of
the standard Wilsonian formulation of lattice gauge theories.

We anticipate that our modified lattice Landau gauge, 
combined with the coset space gauge-fixing as proposed by Schaden,
may turn out to be the most viable candidate to evade the Neuberger $0/0$
problem which has prohibited realizing the 
BRST symmetry on the lattice for~over~25~years.

\section*{Acknowledgment}

The first three authors were supported in part by a DARPA Young Faculty
Award (YFA). NSD and JDH were supported in part by NCSU Faculty Research and
Development Fund, and JDH was supported in part by NSF grant DMS-1262428.
CS was supported by a Rhodes College Faculty Development Endowment grant
and the Ellett Professorship in Mathematics.
A part of this paper is based on DM's thesis, and he would like to thank
Lorenz von Smekal for numerous discussions related to this work. DM would also like to thank Maarten Golterman, Axel Maas, 
Yigal Shamir, Jon-Ivar Skullerud, Andre Sternbeck and Anthony Williams for valuable
discussions related to this work.

\appendix
\section{Results from Homotopy Continuation for the Stereographic Projection}
\label{app:stereographic_LLG_alg_geom}

The following shows that $Z_{GF}$ for the stereographic lattice Landau gauge-fixing functional is orbit independent over the orbit space
except for regions having measure zero.
For this, we first note that the Hessian matrix of Eq.~(\ref{eq:mllg-functional}) is generically positive-definite \cite{Mehta:2009,Hughes:2012hg}.
Hence, $Z_{GF}$ in
Eq.~(\ref{eq:Z_GF_is_Euler_char}) computes the number of stationary points of Eq.~(\ref{eq:mllg-functional}) for the given gauge-orbit.
Thus, we need to compute the number of solutions of Eq.~(\ref{eq:any_dim_mllg_eq}) for various orbits (i.e., random values of
$\{\phi_{\bold{j},\mu}\}$, at the strong coupling limit $\beta=0$) and determine if they remain constant. Finding all the solutions of
such a nonlinear system of equations is a very difficult task. In Refs.\cite{Mehta:2009,Mehta:2009zv,Hughes:2012hg} the problem of solving gauge-fixing conditions
on the lattice was translated in terms of algebraic geometry in order to be able to use the numerical algebraic geometry methods to find
all the solutions of these equations.
With the improved version of the corresponding algorithms, we can now solve the equations for at least the simplest nontrivial lattices in 2D
successfully. To use this method for our purposes, we begin by transforming
our system of trigonometric equations into a system of polynomial equations
by first expanding Eq. (\ref{eq:any_dim_mllg_eq}) using the
trigonometric identity
\begin{equation}
\tan\frac{x+y+z}{2} = \frac{\sin x+\cos z\sin y+\cos y\sin z}{\cos x+\cos y\cos z-\sin y\sin z}.
\end{equation}
Replacing $\sin\theta_{\bold{j}}$ and $\cos\theta_{\bold{j}}$
with $s_{\bold{j}}$ and $c_{\bold{j}}$, resp., yields
\begin{eqnarray}\label{eq:f}
f_{\bold{j}}^{s}(c,s) &=&\sum_{\mu}\Big(\frac{\sin\phi_{\bold{j},\mu}c_{\bold{j}} -\cos\phi_{\bold{j},\mu}s_{\bold{j}}+s_{\bold{j}+\muhat}}{\sin\phi_{\bold{j},\mu}s_{\bold{j}}
+\cos\phi_{\bold{j},\mu}c_{\bold{j}}+c_{\bold{j}+\muhat}}\nonumber \\
& & -\frac{\sin\phi_{\bold{j}-\muhat,\mu}c_{\bold{j}-\muhat}-\cos\phi_{\bold{j}-\muhat,\mu}s_{\bold{j}-\muhat}+s_{\bold{j}}}{\sin\phi_{\bold{j}-\muhat,\mu}s_{\bold{j}-\muhat}+\cos\phi_{\bold{j}-\muhat,\mu}c_{\bold{j}-\muhat}+c_{\bold{j}}}\Big).
\end{eqnarray}
Due to the Pythagorean identity, we add the additional constraint equations
$c_{\bold{j}}^2+s_{\bold{j}}^2-1=0$ for each $\bold{j}$.
As the simplest non-trivial case, we take the $3\times 3$ lattice.
To make sure that there are only isolated solutions, we also fix $\theta_{3,3}=0$
and then remove the equation $f_{3,3}=0$ from the system.
Since the above equations have denominators,
we introduce auxiliary variables
to produce polynomial conditions to satisfy the system. For the $3\times 3$ lattice,
this produces a system of $48$ quadratic polynomial equations in $48$ variables
that depends on 18 parameters $\{\phi_{\bf j,\mu}\}$.
This procedure is a one-to-one transformation so that
no solutions of the original system are lost in the transformation.

\subsection{Methods}

We solve the system consisting of $48$ equations using a two-phase
methodology from numerical algebraic geometry known as a \emph{parameter
homotopy} which guarantees to find \text{all} the solutions
of a given system of multivariate polynomial equations for any given
parameter points. We give a brief summary; for further details, see
Refs. \cite{SW:95,BHSW13} and Refs.~\cite{Mehta:2011wj,Mehta:2009zv,Maniatis:2012ex,
Kastner:2011zz,Mehta:2012wk,Mehta:2012qr,Nerattini:2012pi,Hughes:2012hg,Hauenstein:2012xs,
Greene:2013ida,Mehta:2013fza,MartinezPedrera:2012rs,He:2013yk} for the related applications in lattice field theories
and other particle physics areas.

First, in the \emph{ab initio} phase, we choose a random set of complex
parameters $P_0:=\{\phi^*_{\bold{j},\mu}\}$ and numerically
compute the set of solutions $S_0$ to the system using homotopy continuation with
regeneration \cite{HSW11}, implemented in
{\tt Bertini} \cite{BHSW06}.  This phase, which is performed only once,
took roughly $20.5$ hours on a cluster consisting of
four AMD 6376 Opteron processors, i.e., $64$ computing
cores running at 2.3 GHz.  Subsequent computations make
use of these results to significantly reduce the effort involved in
solving the system.  In particular, we find that there are $11664$
nonsingular isolated solutions for the random set of parameters $P_0$.

In the second phase, known as the \emph{parameter homotopy phase}, we
solve the system for various choices of parameters.  For each
set of parameters $\{\phi_{\bold{j},\mu}\}$, we use {\tt Bertini} to
numerically track paths starting at the points
in $S_0$.  We numerically follow paths defined by a continuous
deformation of the parameters from $P_0$ to $\{\phi_{\bold{j},\mu}\}$, so
that the endpoints are the solutions we seek.
On the same cluster,
this phase takes an average of $39$ minutes to
compute solutions for a given set of parameters.\\

\subsection{Results}

First, to determine the behavior of the system at general points in the
parameter space, we solved the system for $780$ random sets of
real parameters $\{\phi_{\bold{j},\mu}\}$.  In each instance, we find that there are
$11664$ real solutions.  Thus, we conjecture that all
$11664$ complex solutions are real for all points in the real parameter
space except on several regions.

Next, we investigate the discriminant locus, which is the set on which the system has nongeneric behavior.  We find that when
the angles in $\{\phi_{\bold{j},\mu}\}$ are deliberately chosen so that they
adhere to some structure, such as rational multiples of $\pi$, it is
quite easy to find a point in the parameter space such that the system
has fewer than $11664$ real solutions.  Thus, the number of stationary
points of Eq.~(\ref{eq:mllg-functional}) differs for various orbits,
and $Z_{GF}$ for the stereographic lattice Landau gauge-fixing
functional is orbit-dependent.

\begin{figure}[!ht]
  \centering
  \includegraphics[width=0.36\textwidth]{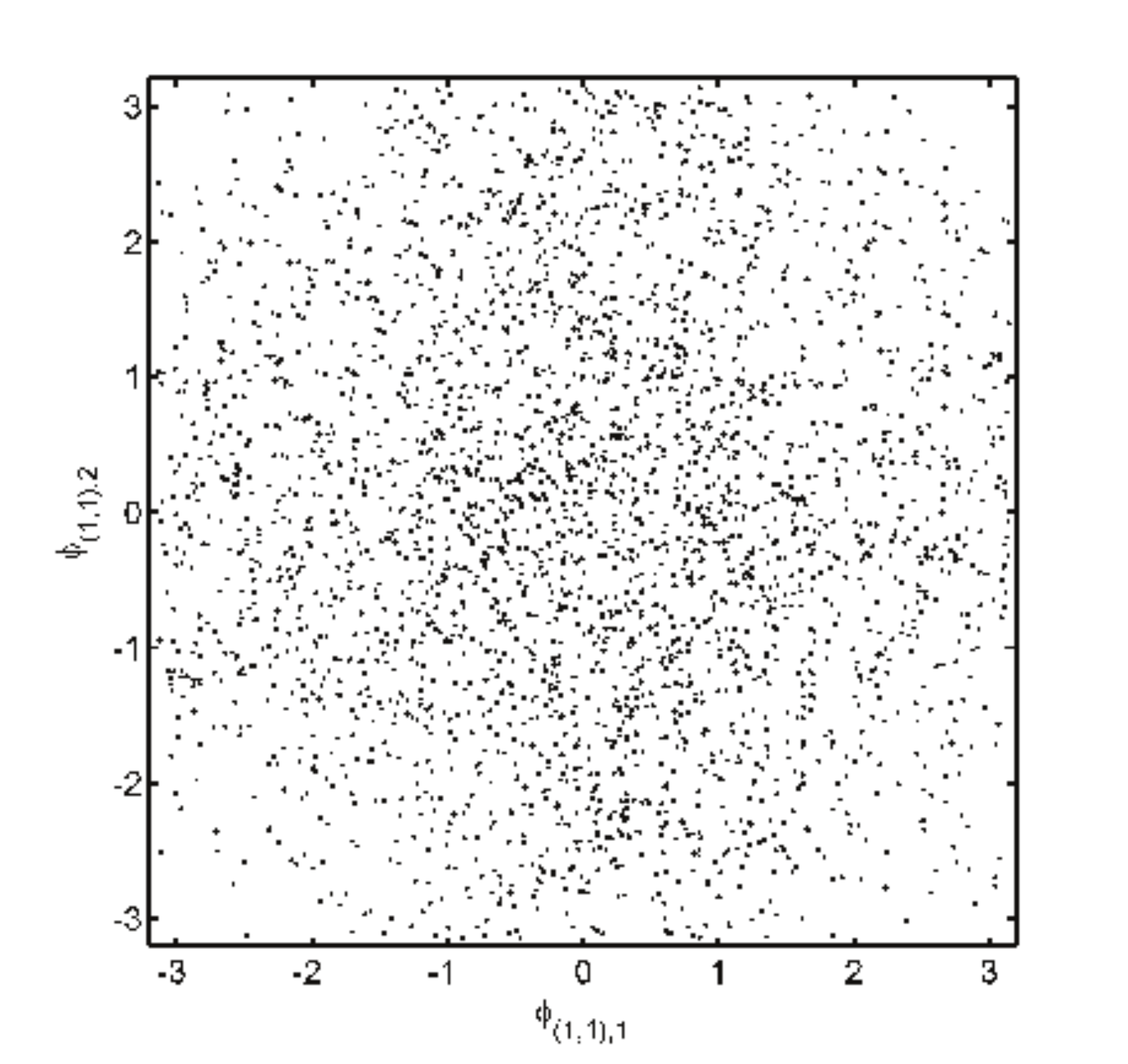}
  \caption{Subset of the discriminant locus projected onto two parameters for the $3\times3$ lattice. The other $\phi$s are fixed as listed in
  Table \ref{table:FixedParameterValues}. While varying $\phi_{(1,1),1}$ and $\phi_{(1,1),2}$ and leaving all other $\phi$s fixed for the $3\times 3$
  lattice, the points in this plot are the points at which $Z_{GF}$ differs from the generic value $11664$.}
\label{figure:DiscriminantLocus}
\end{figure}

The following figures summarize these results.
Figure~\ref{figure:SpecialParameters} plots $Z_{GF}$ (or, equivalently,
the number of real solutions) corresponding to various sets of parameters
$P_1,\ldots,P_4$.  Figure \ref{figure:DiscriminantLocus} plots a
subset of the discriminant locus projected onto the two parameters
$\phi_{(1,1),1}$ and $\phi_{(1,1),2}$ in which the rest of the parameters
are fixed to the angles given in
Table \ref{table:FixedParameterValues}.  To locate points on the
discriminant locus, we used the fact that for parameter values to have
fewer than $11664$ real solutions, we must have corresponding denominators
equal to zero in Eq.~(\ref{eq:f}).  Since we introduced auxiliary variables
for denominators when constructing the polynomial system, we can perform
parameter homotopies in which the destination systems have
these `denominators' equal to zero.  We note that the points shown here are
only a subset of the discriminant locus, which is an algebraic curve in
this projection.  Nevertheless, these computed points illustrate the
abundance of parameter choices for which the system has nongeneric behavior.

\begin{figure}[!ht]
  \centering
  \includegraphics[width=0.38\textwidth]{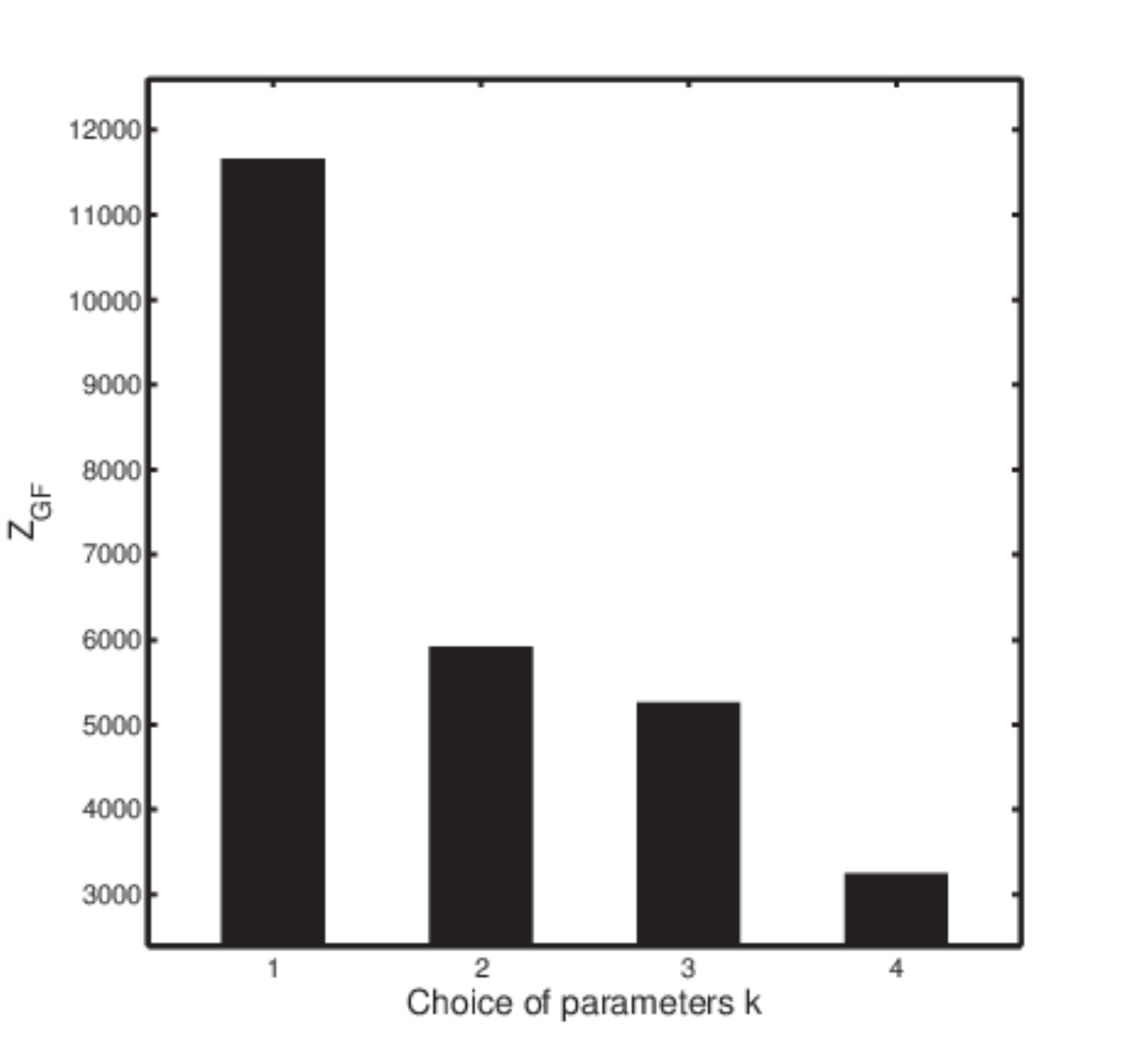}
  \caption{$Z_{GF}$ corresponding to various sets of parameters $P_k$ which are defined as follows.
           For $P_1$, we set each parameter to a distinct angle via
           $\phi_{\bold{j},\mu} = \pi/(j_2+3(j_1-1)+9(\mu-1))$.
           For $P_2$, we set $\phi_{\bold{j},\mu}=\pi /2$ for all $\bold{j}$ and $\mu$.
           For $P_3$, we set $\phi_{\bold{j},\mu}=0$ for all $\bold{j}$ and $\mu$.
           For $P_4$, we set $\phi_{\bold{j},\mu}=\pi/3 + (\pi/6)(\mu -1).$
          }
\label{figure:SpecialParameters}
\end{figure}

\begin{table}[!ht]
\centering
{\scriptsize
\begin{tabular}{|c|c|c|c|c|c|c|c|c|c|c|c|c|c|c|c|c|}
  \hline
  $j_1$ & $1$ & $1$ & $2$ & $2$ & $2$ & $3$ & $3$ & $3$ \\
  \hline
  $j_2$ & $2$ & $3$ & $1$ & $2$ & $3$ & $1$ & $2$ & $3$ \\
  \hline
  $\mu$ & $1$ & $1$ & $1$ & $1$ & $1$ & $1$ & $1$ & $1$ \\
  \hline
  $\phi_{(j_1,j_2),\mu}$ &  $-\frac{\pi}{2}$ & $\frac{\pi}{5}$ & $-\frac{5\pi}{11}$
  & $\frac{15\pi}{17}$ & $-\frac{15\pi}{23}$ & $\frac{28\pi}{31}$
  & $\frac{24\pi}{41}$ & $-\frac{7\pi}{47}$\\ [1ex]
  \hline
\end{tabular}
\\\ \\\ \\
\begin{tabular}{|c|c|c|c|c|c|c|c|c|c|c|c|c|c|c|c|c|}
  \hline
  $j_1$ & $1$ & $1$ & $2$ & $2$ & $2$ & $3$ & $3$ & $3$ \\
  \hline
  $j_2$ & $2$ & $3$ & $1$ & $2$ & $3$ & $1$ & $2$ & $3$ \\
  \hline
  $\mu$ & $2$ & $2$ & $2$ & $2$ & $2$ & $2$ & $2$ & $2$ \\
  \hline
  $\phi_{(j_1,j_2),\mu}$ & $\frac{2\pi}{3}$ & $-\frac{5\pi}{7}$
  & $\frac{\pi}{13}$ & $\frac{17\pi}{19}$ & $\frac{27\pi}{29}$
  & $-\frac{\pi}{37}$ & $-\frac{30\pi}{43}$ & $\frac{44\pi}{53}$\\ [1ex]
  \hline
\end{tabular}
}
\caption{Fixed parameter values used for Figure \ref{figure:DiscriminantLocus}.}
\label{table:FixedParameterValues}
\end{table}

\newpage

\bibliographystyle{unsrt}
\bibliography{bibliography}

\end{document}